\documentclass{Interspeech}

\interspeechcameraready

\title{MIKU-PAL: An Automated and Standardized Multimodal Method for \\ Speech Paralinguistic and Affect Labeling}

\author[affiliation={1,3}]{Yifan}{Cheng}
\author[affiliation={1,4}]{Ruoyi}{Zhang}
\author[affiliation={2}]{Jiatong}{Shi}

\affiliation{}{Fish Audio}{Santa Clara, CA, USA}
\affiliation{}{Carnegie Mellon University}{Pittsburgh, PA, USA}
\affiliation{}{Huazhong University of Science and Technology}{Wuhan, Hubei, China}
\affiliation{}{Nanjing University of Information Science and Technology}{Nanjing, Jiangsu, China}
\email{yf\_cheng@hust.edu.cn, potato\_zhang@nuist.edu.cn, jiatongs@cs.cmu.edu}

\usepackage{comment}
\usepackage{subcaption}
\usepackage{booktabs}
\usepackage{multirow}
\usepackage{array}
\usepackage{ragged2e}
\usepackage{svg}
\usepackage{graphicx}
\usepackage{cite}

\begin{document}

\maketitle

\begin{abstract}
    Acquiring large-scale emotional speech data with strong consistency remains a challenge for speech synthesis. This paper presents MIKU-PAL, a fully automated multimodal pipeline for extracting high-consistency emotional speech from unlabeled video data. Leveraging face detection and tracking algorithms, we developed an automatic emotion analysis system using a multimodal large language model~(MLLM). Our results demonstrate that MIKU-PAL can achieve human-level accuracy (68.5\% on MELD) and superior consistency (0.93 Fleiss $\kappa$ score) while being much cheaper and faster than human annotation. With the high-quality, flexible, and consistent annotation from MIKU-PAL, we can annotate fine-grained speech emotion categories of up to 26 types, validated by human annotators with 83\% rationality ratings. Based on our proposed system, we further released a fine-grained emotional speech dataset MIKU-EmoBench(131.2 hours) as a new benchmark for emotional text-to-speech and visual voice cloning.\end{abstract}

\section{Introduction}

Modern human-computer interaction systems increasingly demand emotional speech synthesis capabilities. However, a fundamental bottleneck exists: data scarcity for emotional speech. While current speech language models are usually trained on millions of hours of data~\cite{du2024cosyvoice, anastassiou2024seed, wang2023neural}, emotional speech datasets remain limited.  Existing resources like IEMOCAP~\cite{busso2008iemocap}, MELD~\cite{poria2019meld}, and MSP-Podcast~\cite{lotfian2017building} are manually annotated, resulting in datasets with typically less than 300 hours. Meanwhile, the current speech emotion datasets are constrained to only 5-8 emotion categories, often following Ekman's basic emotion framework~\cite{ekman1992facial}. These categories are increasingly being challenged by psychological research~\cite{sabini2005ekman, barrett2006emotions, cowen2017self}, which suggests a richer and more nuanced spectrum of human emotions far beyond these basic sets. This lack of emotional richness in speech datasets is further emphasized when compared to Natural Language Processing~(NLP), where datasets feature up to 27 emotion categories~\cite{demszky2020goemotions}. 

The core problem is the manual annotation process, burdened by high costs and low consistency. Datasets like EmoDB~\cite{burkhardt2005database} and RAVDESS~\cite{livingstone2018ryerson} using acted emotions and later datasets like IEMOCAP with improvised scenarios, and MELD and MSP-Podcast aiming for naturalness, all ultimately rely on this expensive and inconsistent manual labeling.  This limits both the size and the emotional diversity of available datasets, hindering the realization of fine-grained emotion modeling, especially to meet the growing need for downstream applications like emotional text-to-speech (TTS).
While recent efforts explore learned emotion representations to bypass direct labels~\cite{wu2024laugh, tang2024ed}, they have to sacrifice explicit user control over synthesized affect.

To overcome these limitations, we introduce \textbf{MIKU-PAL}~(Multimodal Intelligence Kit for Understanding - Paralinguistic and Affect Labeling), a novel multimodal framework designed to automate emotion annotation across audio, visual, and text modalities. MIKU-PAL achieves high-consistency emotion judgments (Fleiss $\kappa$ 0.93) with flexible emotion categories. With the high-quality, flexible, and consistent annotation from MIKU-PAL, we expand emotion categories to 26, drawing from psychological research~\cite{cowen2017self}, validated through human rationality ratings, and downstream TTS applications. With low cost and high efficiency, MIKU-PAL achieves 63.6\% accuracy with released labels on IEMOCAP and MELD. The system also achieved typical human annotator accuracy\footnote{This was obtained by analyzing the raw annotation data from each annotator in the IEMOCAP dataset.} on the datasets. With MIKU-PAL, we created MIKU-EmoBench (131.2 hours, 26 emotions). It outperforms the original emotional speech dataset in a variety of metrics, as shown in Table~\ref{tab:miku_emobench_performance}.


\begin{figure}
    \centering
    \includegraphics[width=\linewidth]{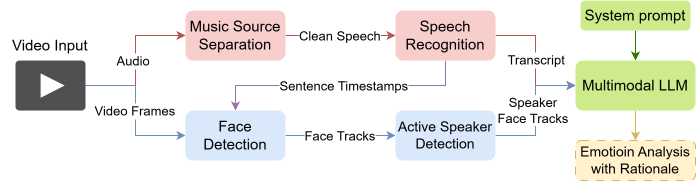}
    \caption{The structure overview of MIKU-PAL: It analyzes visual, text, and audio modalities across three stages.}
    \vspace{-0.5cm}
    \label{fig:pipeline}
\end{figure}

{\raggedright
In summary, our main contributions are as follows:
\par}

\begin{itemize}[leftmargin=2.5mm]
    \item We develop \textbf{MIKU-PAL}, the first multimodal automated emotion labeling system, capable of automatically annotating high-quality, flexible, and consistent emotion labels at low cost.
    \item Based on psychological research and TTS performance, we investigated a standardized speech emotion classification system consisting of 26 categories.
    \item Release an open-source multimodal emotional speech dataset called \textbf{MIKU-EmoBench} with the new classification, to serve as a new benchmark for emotional speech tasks.
\end{itemize}

\section{MIKU-PAL}

In this section, we introduce the detailed design of MIKU-PAL. First, we discuss the general pipeline in Sec.~\ref{pipeline}. Then we discuss the design of emotion categories in Sec.~\ref{emotion}. In Sec.~\ref{performance}, we present a thorough evaluation of the data by checking the annotation quality with human annotators.

\subsection{Automatic Pipeline}\label{pipeline}

Our pipeline comprises three stages as shown in Fig.~\ref{fig:pipeline}: audio preprocessing, vision preprocessing, and emotion analysis. MIKU-PAL is a modularized pipeline where each component can be separately upgraded to improve the performance of the whole framework.

\begin{figure}
    \centering
    \includegraphics[width=0.9\linewidth]{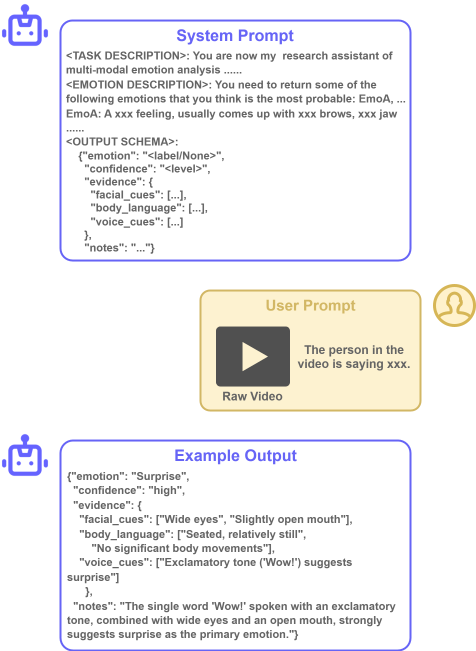}
    \caption{Overview of the MLLM chat. The system prompt is based on three parts: mission description, textual description of emotions, and output structure. The user prompt only contains raw video and text. Example output presents a representative example of the system output.}
    \vspace{-0.5cm}
    \label{fig:gemini_prompt}
\end{figure}

\noindent \textbf{Audio Preprocessing}. Raw audio in multimodal emotion analysis often contains background music and environmental noise, degrading emotional speech segment quality. To purify audio, our pipeline uses MDX-Net~\cite{kim2021kuielab} as a Music Source Separation~(MSS) model to extract vocals. MDX-Net is a dual-stream architecture designed to isolate clean vocal signals from complex audio mixtures. It was chosen for its state-of-the-art performance in MSS.~\cite{solovyev2023benchmarks}. Analysis of 30 Friends episodes shows a 36\% improvement in Signal-to-Noise Ratio (SNR) after separating speech and non-speech segments using WebRTC Voice Activity Detection (VAD). Subsequently, to obtain transcriptions for temporal alignment, we utilized the Whisper-large v3 model~\cite{radford2023robust}. It provides timestamps for vision preprocessing and transcriptions for subsequent MLLM emotion analysis.

\noindent \textbf{Vision Preprocessing}. Audio-processed speech segments proceed to face detection, employing S$^3$FD~\cite{zhang2017s3fd} or DSFD~\cite{li2019dsfd} based on computational needs.  S$^3$FD's efficient, scale-invariant architecture excels in multi-scale face detection, while DSFD's deeper network offers higher accuracy, especially with pose variations or occlusions. In our task, speakers are typically centered and prominent in our analyzed video frames, thus prioritizing processing speed over maximal accuracy.

For effective audio-visual fusion, active speaker identification is crucial.  MIKU-PAL integrates TalkNet~\cite{tao2021someone} for this task, leveraging its 92\% accuracy on the AVA Active Speaker Detection challenge, ensuring robust performance. Accurate speaker identification allows precise attribution of facial and vocal emotion features, enhancing MLLM-based multimodal analysis.

In addition, we conducted an ablation study to assess the impact of visual preprocessing on emotion analysis. Evaluation on the MELD test set showed a 25.6\% improvement in annotation accuracy when comparing results with and without visual preprocessing. This highlights visual information's significant role in MIKU-PAL.

\noindent \textbf{MLLM Emotion Analysis}. MLLMs have shown promise in emotion analysis~\cite{cheng2024emotion, lin2024moe}. Based on this, MIKU-PAL employs Gemini 2.0 Flash~\cite{team2023gemini} for emotion analysis using facial, audio, and textual features. Our prompt design is illustrated in Fig.~\ref{fig:gemini_prompt}. The system prompt comprises three key components: a task description, a textual emotion description, and an output structure. The output structure is designed to guide the MLLM in assessing emotion characteristics across different modalities and ultimately providing a rationale in natural language for its judgment. The user prompt contains only raw video and text. This zero-shot approach allows MIKU-PAL to conduct flexible emotion classification with consistent criteria. 

\subsection{Emotion Categories}\label{emotion}

\begin{figure}
    \centering
    \includegraphics[width=7.5cm]{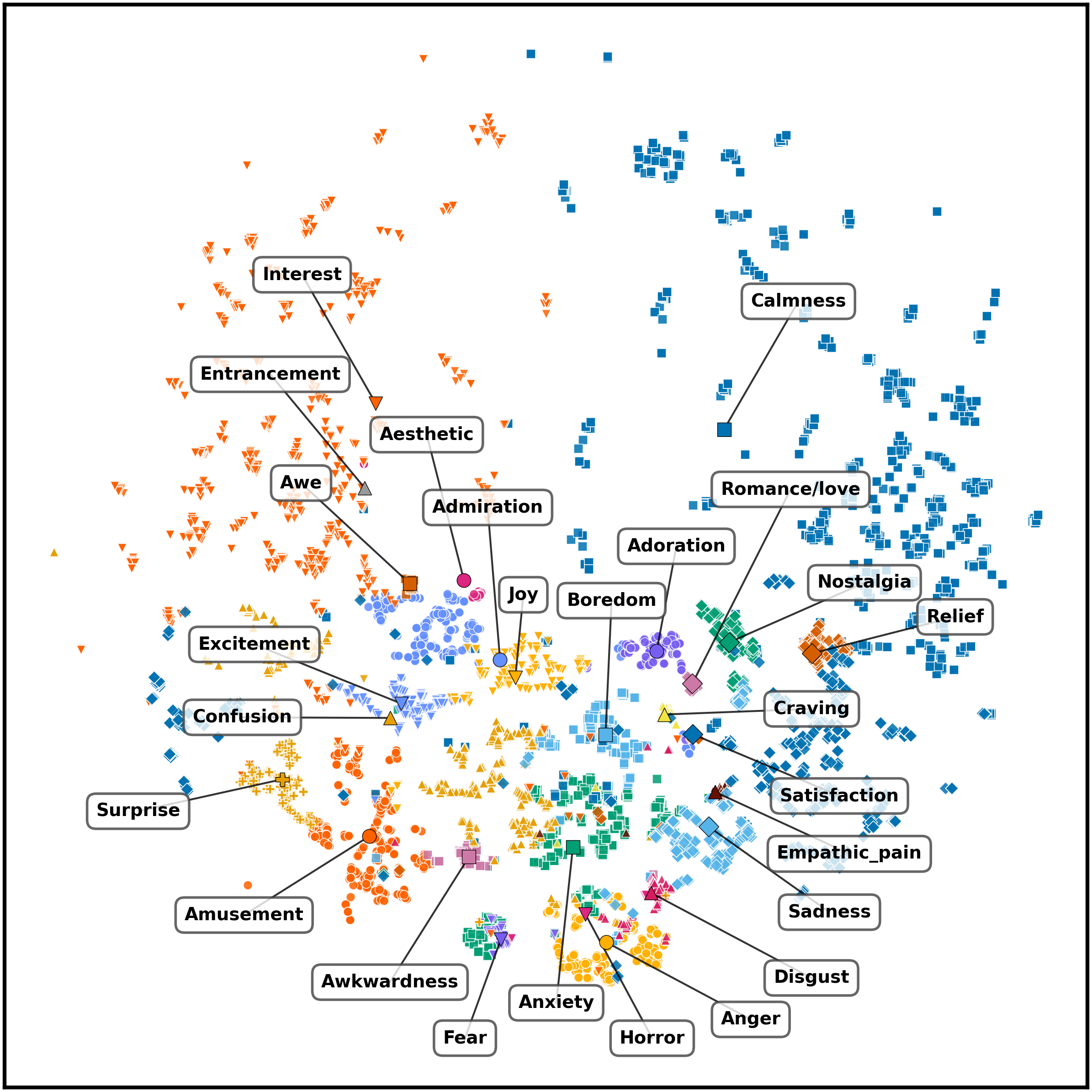}
    \caption {Mixed emotion analysis on 10,000 YouTube video segments using MIKU-PAL. Annotation results are reduced to two dimensions using t-SNE. Each data point is labeled according to the emotion category with the highest intensity and colored using a weighted interpolation based on all emotion categories present. The visualization demonstrates MIKU-PAL's ability to model the continuous human emotion space and the gradient relationships between emotion categories.}
    \vspace{-1em}
    \label{fig:emotion_matrix}
\end{figure}

Anger, disgust, fear, joy, sadness, and surprise, as basic emotion categories, have been extensively utilized in emotional speech datasets, such as EmoDB~\cite{burkhardt2005database}, RAVDESS~\cite{livingstone2018ryerson}, and MELD \cite{poria2019meld}. These categories have seen widespread adoption in Speech Emotion Recognition (SER) tasks~\cite{kounadis2024wav2small, ma2023emotion2vec}. However, recent research indicates that these limited emotion categories do not meet the growing needs of various downstream tasks, such as emotional TTS~\cite{cho2024emosphere++, kuligowska2018speech}. Contemporary psychological research, exemplified by Cowen et al.’s work~\cite{cowen2017self} proposes a more comprehensive model: human emotions exist as complex mixtures within a continuous space defined by 27 discrete categories.\footnote{These categories include: Admiration, Adoration, Aesthetic, Amusement, Anger, Anxiety, Awe, Awkwardness, Boredom, Calmness, Confusion, Craving, Disgust, Empathic pain, Entrancement, Excitement, Fear, Horror, Interest, Joy, Romance/Love, Nostalgia, Relief, Sadness, Satisfaction, Surprise. Due to ethical considerations, one category was removed from the original set.} We construct our pipeline's emotion categories upon this richer, psychologically validated framework, aiming for a more nuanced and expressive emotion representation.

To validate MIKU-PAL's capability to capture these 26 emotion categories, we conducted a mixed emotion annotation experiment on 10,000 randomly collected YouTube video segments. The annotation process was consistent with the standard MIKU-PAL pipeline. The key difference was the incorporation of Cowen's emotion categories paper as an additional prompt. All emotion annotation results were then reduced to two dimensions using t-distributed stochastic neighbor embedding (t-SNE) and visualized in Fig.~\ref{fig:emotion_matrix}.

This map reveals the trajectory and distribution of human emotions, illustrating transitions such as from admiration to love, and from joy to satisfied to excitement.  These observed patterns are largely consistent with the conclusions of the original psychology research paper, demonstrating the rationality of MIKU-PAL in capturing this expanded emotion categorization.  Meanwhile, to investigate whether MIKU-PAL's annotations align with human perceptions, we conducted a human rationality annotation experiment. We recruited 5 annotators without relevant background to evaluate 1000 balanced samples across emotion categories, assessing the reasonableness of MIKU-PAL's annotations. The results showed that human annotators considered 83\% of MIKU-PAL's annotations to be reliable. 

\begin{figure}[!t]
    \centering
    \includegraphics[width=3.7cm]{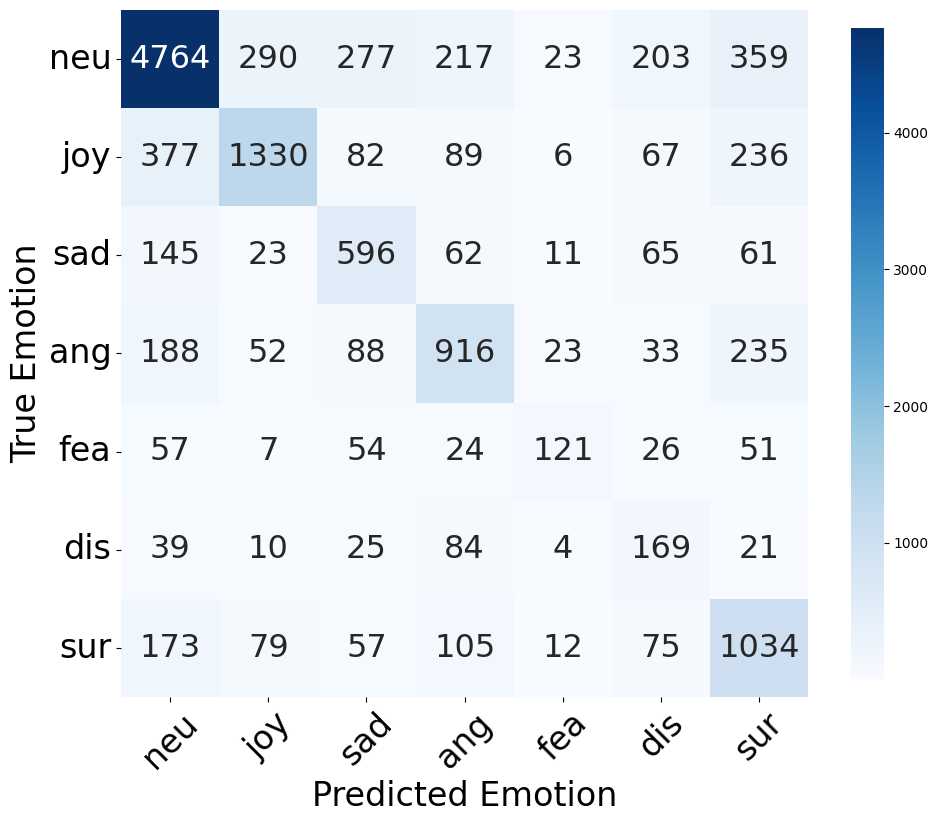}
    \hspace{0.15cm}
   \includegraphics[width=3.7cm]{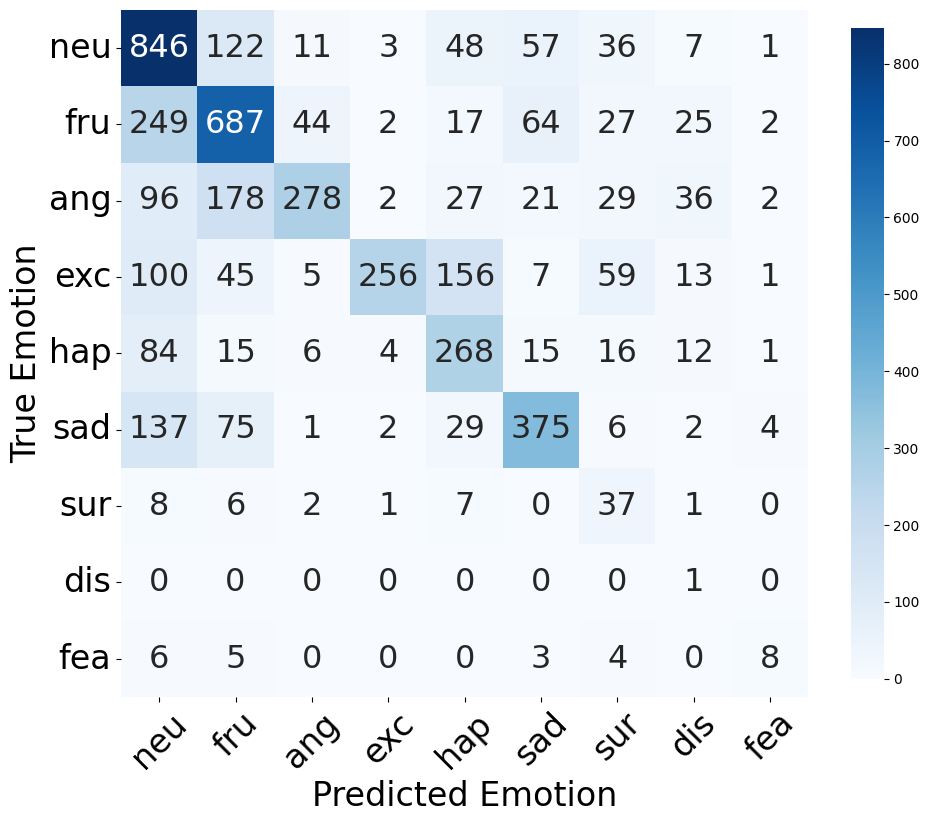}
    \caption{Confusion matrix of MELD (left) and IEMOCAP (right). It demonstrates good performance in emotions that have been psychologically validated.}
    \vspace{-1em}
    \label{fig:confusion_matrix}
\end{figure}

\subsection{Performance} \label{performance}
To comprehensively evaluate the effectiveness of MIKU-PAL, we conducted a series of experiments. Specifically, our evaluation focuses on the following key aspects: engineering performance, accuracy, consistency, and emotional TTS performance. The core performance metrics compared to typical human annotators are shown in Table \ref{tab:miku_pal_performance}.

\begin{table}[]
    \centering
        \caption{The performance of MIKU-PAL compared to typical human annotators.}
    \scalebox{0.85}{
    \begin{tabular}{lccccc}
        \toprule
        \multirow{2}{*}{Dataset} & \multirow{2}{*}{Method} & Acc & \multirow{2}{*}{Fleiss $\kappa$} & Cost & Relative \\
        & & (\%) & & (/hour) & MOS \\
        \midrule
        \multirow{2}{*}{IEMOCAP}  
        & Human & 72.9 & 0.40 & $\sim$10\$ & \multirow{2}{*}{+0.07} \\
        & MIKU-PAL & 58.6 & \textbf{0.93} & \textbf{\textless 50\textcent}  \\
        
        \addlinespace[0.5em]
        \multirow{2}{*}{MELD} 
        & Human & N/A & 0.43 & $\sim$10\$ & \multirow{2}{*}{+0.09}\\
        & MIKU-PAL & 68.5 & \textbf{0.95} & \textbf{\textless 50\textcent}  \\
        \bottomrule
    \end{tabular}}

    \vspace{-1.5em}
    \label{tab:miku_pal_performance}
\end{table}

\noindent \textbf{Engineering}. From an engineering perspective, MIKU-PAL demonstrates significant advantages in terms of both processing speed and cost-effectiveness. We evaluated the pipeline's performance on a standard workstation equipped with 8 NVIDIA RTX 4090 GPUs. MIKU-PAL achieves a processing speed ratio of approximately 1:12 on 720p 30fps video data. Furthermore, it maintains a retention rate of 42\% when processing randomly collected YouTube videos. This means that processing 100 hours of raw videos yields 42 hours of final audio. We have implemented parallel processing for the entire pipeline. GPU processing speed remains the primary bottleneck. MIKU-PAL exhibits near-lossless performance under multi-threading parallelism and allows for the independent specification of parallelism for each stage, ensuring efficient utilization of computational resources. Utilizing the latest Google Gemini 2.0 Flash model, the cost of generating 1 hour of emotional speech dataset data using MIKU-PAL is approximately 50 US cents. This is substantially lower than the cost associated with manual annotation.

\begin{table*}[ht]
\centering
\caption{Multimodal Emotion Dataset Characteristics Comparison}
\label{tab:miku_emobench_performance}
\begin{tabular}{@{}lcccccc>{\centering\arraybackslash}p{1.1cm}@{}}
\toprule
\multirow{2}{*}[-\dimexpr\baselineskip/4]{Dataset} & 
\multirow{2}{*}[-\dimexpr\baselineskip/4]{Visual} & 
\multirow{2}{*}[-\dimexpr\baselineskip/4]{Domain} & 
Duration & Segments & Emotion & Labeling & 
\multirow{2}{*}[-\dimexpr\baselineskip/20]{Updates} \\
& & & (h) & (k) & Categories & Method & \\
\midrule
IEMOCAP~\cite{busso2008iemocap} & Y & Scripted and spontaneous dialogs & 12.0 & 10.0 & 9 & Human & N \\
EMO-DB~\cite{burkhardt2005database} & N & Studio recorded & 0.5 & 0.5 & 7 & Preconfigured & N \\
MELD~\cite{poria2019meld} & Y & TV show & 13.7 & 13.7 & 7 & Human & N \\
CMU-MOSEI~\cite{zadeh2018multimodal} & Y & YouTube opinion videos & 65.9 & 23.4 & 6 & Human & N \\
MSP-podcast~\cite{lotfian2017building} & N & Podcast & \textbf{237.9} & \textbf{151.6} & 7 & Human & Y \\
\midrule[0.8pt]
\textbf{MIKU-EmoBench} & \textbf{Y} & \textbf{Almost all domains on Youtube} & 131.2 & 66.0 & \textbf{26} & \textbf{Automatic} & \textbf{Y}\\
\bottomrule
\end{tabular}
\end{table*}

\begin{table*}
\centering
\caption{Characteristics Comparison of Emotional Speech Datasets with MIKU-EmoBench}
\label{tab:miku_emobench_benchmark}
\begin{tabular}{@{}lccccc@{}}
\toprule
\multirow{2}{*}{Model} & WER(\%)  & SS & MOS & Emotion Similarity \\
& $\downarrow$ & $\uparrow$ & $\uparrow$ & $\uparrow$ \\
\midrule
Fish-Speech~\cite{liao2024fish} & \textbf{2.4} & 0.762 & 3.91 & 0.88 \\
IEMOCAP-ft Fish-Speech & 2.6 & 0.780 & 4.01 & 0.89 \\
MELD-ft Fish-Speech & 2.5 & 0.775 & 4.00 & 0.89 \\
MSP-Podcast-ft Fish-Speech & \textbf{2.4} & 0.785 & 4.09 & 0.91 \\
CosyVoice~\cite{du2024cosyvoice} & 2.6 & \textbf{0.802} & 3.97 & 0.87 \\
\midrule[0.8pt]
\textbf{MIKU-EmoBench-ft Fish-Speech} & \textbf{2.4} & 0.792 & \textbf{4.12} & \textbf{0.92} \\
\bottomrule
\end{tabular}
\vspace{-1em}
\end{table*}

\noindent \textbf{Accuracy}. We validated the accuracy of our pipeline on the IEMOCAP and MELD datasets. The results indicate that we achieved an overall accuracy of approximately 65\%. As depicted in the confusion matrix in Fig.~\ref{fig:confusion_matrix}, wrong classifications are notably frequent between 'frustration' and 'neutral' emotions.  Interestingly, these two emotion categories are not recognized as distinct basic emotions within established psychological emotion classifications; rather, they are often considered to be encompassed within other broader emotional categories. When these two emotions are excluded, MIKU-PAL's accuracy reaches approximately 75\%, which surpasses the average accuracy of human annotators. This result further substantiates the effectiveness of our pipeline and highlights potential limitations in current emotion classification schemes. Confusion matrix analysis reveals classification overlap between 'excited' and 'happy' within the IEMOCAP dataset. This confusion mainly arises because IEMOCAP assigns one label as final when emotions receive equal annotation counts.

\noindent \textbf{Consistency}. To evaluate the annotation consistency of MIKU-PAL, we calculated the Fleiss' Kappa score \cite{fleiss1971measuring} on the IEMOCAP and MELD datasets. The Fleiss' Kappa score is a statistical measure used to assess the agreement among multiple annotators, with higher values indicating greater consistency in annotation. Across five independent annotation experiments, we maintained a consistent prompt and model configuration. MIKU-PAL achieved Fleiss' Kappa scores of 0.93 on IEMOCAP and 0.95 on MELD. The result indicates a very high level of annotation consistency significantly surpassing the consistency levels typically attained by human annotators.

\noindent \textbf{Emotional TTS Performance}. To validate the effectiveness of MIKU-PAL for the targeted emotional TTS task, we re-annotated IEMOCAP and MELD datasets with MIKU-PAL. And used the re-annotated dataset with special emotion tokens to fine-tune Fish-Speech~\cite{liao2024fish}. We then compared the relative Mean Opinion Score (MOS) of this fine-tuned model against a baseline model fine-tuned on the original datasets. The results indicate that the fine-tuned model achieved a significant MOS score (+0.08) improvement compared to the baseline. Furthermore, both models demonstrated effectiveness in explicit emotion control within TTS. This outcome substantiates the validity of MIKU-PAL annotated data for emotional TTS tasks.

\section{MIKU-EmoBench}

Addressing the critical limitations of existing emotional datasets in data scale and emotion granularity for next-generation emotional TTS, we developed MIKU-EmoBench\footnote{Dataset avaliable at https://huggingface.co/datasets/WhaleDolphin/MIKU-EmoBench}, a novel dataset collected from YouTube videos using our MIKU-PAL pipeline. MIKU-PAL's inherent efficiency directly enabled the rapid collection of 131.2 hours of data within a single week, a collection speed that significantly surpasses any existing emotional speech dataset. Furthermore, leveraging MIKU-PAL's fine-grained emotion analysis capabilities, MIKU-PAL has 26 psychologically proven emotion categories better suited for the nuanced demands of next-generation emotional TTS systems. As Table \ref{tab:miku_emobench_performance} shows, MIKU-EmoBench outperforms existing datasets across multiple metrics. Meanwhile, the annotation files will be made publicly available and continuously updated to facilitate downstream emotional speech tasks.

\vspace{-0.6em}
\subsection{Data Information}
Our MIKU-EmoBench dataset comprises 131.2 hours of emotion-labeled audio, segmented into 65,970 utterances with an average duration of 7.16 seconds (min. 2s). To ensure diversity, MIKU-EmoBench incorporates audio from various scenes (e.g., interviews, movies, daily conversations), countries and regions (e.g., USA, Europe, Asia) and races (e.g., Caucasian, Asian, African descent).  This rich diversity, sourced from open YouTube videos,  provides a broad representation of emotional speech.  Annotations cover 26 mixed emotions, each with intensity scores and textual rationales, offering detailed and nuanced emotional information.

\vspace{-0.6em}
\subsection{Baseline experiments}
For baseline evaluation and to best demonstrate MIKU-EmoBench's utility, we concentrated our experiments on Emotional TTS. This focus is primarily driven by two factors: first, MIKU-PAL and MIKU-EmoBench are specifically designed for emotional TTS; second, current Speech Emotion Recognition (SER) models are not efficient in discerning the 26 fine-grained emotion categories within our dataset.

We selected Fish-Speech and CosyVoice as our baseline models. We chose Fish-Speech due to its state-of-the-art (SOTA) performance~\cite{tts-arena} and open-source availability. We performed incremental fine-tuning of Fish-Speech using emotional speech datasets, treating emotion labels as special tokens, resulting in a fine-tuned model capable of emotion control via special tokens. CosyVoice, inherently capable of emotion description using natural language, serves as a representative model employing latent variable emotion control.

We fine-tuned Fish-Speech with emotion special tokens using IEMOCAP, MELD (train set), MSP-Podcast, and MIKU-EmoBench, and uniformly tested them using the MELD test set\footnote{Converting MELD's emotional categories to MIKU-EmoBench's available emotional categories for generation.}. Table~\ref{tab:miku_emobench_benchmark} shows the performance metrics, including WER, Speaker Similarity (measured by VERSA~\cite{shi2024versa}), human-annotated MOS, and emotion similarity (computed using FunASR emotion vectors~\cite{ma2023emotion2vec}). Results demonstrate that MIKU-EmoBench fine-tuning improved both MOS and emotion similarity while maintaining TTS quality. This demonstrates the effectiveness of MIKU-PAL and MIKU-EmoBench in this task.

\vspace{-1em}
\section{Conclusion}
In this paper, we present MIKU-PAL, a novel, automated, and flexible multimodal pipeline for emotional speech annotation. MIKU-PAL efficiently and cost-effectively collects emotion datasets with high consistency and human-level accuracy, addressing a field bottleneck. Furthermore, we used MIKU-PAL to develop a large-scale emotion dataset MIKU-EmoBench. This 131-hour YouTube dataset, annotated with 26 fine-grained emotions, showcases MIKU-PAL's potential for generating rich emotion datasets previously infeasible with manual methods. While acknowledging model-dependent performance and potential biases from YouTube data, future work will enhance MIKU-PAL's accuracy, robustness, and adaptability.

\section{Acknowledgement}

Experiments of this work used the Bridges2 system at PSC through allocations CIS210014 and IRI120008P from the Advanced Cyberinfrastructure Coordination Ecosystem: Services \& Support (ACCESS) program, supported by National Science Foundation grants \#2138259,\#tel:2138286, \#tel:2138307, \#tel:2137603, and \#tel:2138296.

\bibliographystyle{IEEEtran}
\bibliography{mybib}

\end{document}